\documentclass[traditabstract]{aa}
\usepackage{graphicx}
\usepackage{amsmath,amssymb}
\usepackage[applemac]{inputenc}
\usepackage{xcolor}
 \usepackage{txfonts}
 \usepackage{natbib}
 \bibpunct{(}{)}{;}{a}{}{,} % to follow the A&A style
\newcommand{\numax}{\nu_\mathrm{max}}
\newcommand{\dnu}{\Delta\nu}
\newcommand{\diamonds}{\textsc{Diamonds}}
\newcommand{\gar}{\textsc{GARSTEC }}
\newcommand{\ctd}{\textsc{CTDYN }}

\def\be{\begin{equation}}
\def\ee{\end{equation}}    
\def\ba{\begin{eqnarray}}
\def\ea{\end{eqnarray}}

\usepackage{color}

\usepackage{epstopdf}

\begin{document}

\title{Acoustic oscillations and dynamo action in the G8 sub-giant EK~Eri\thanks{Based on observations made with the Hertzsprung SONG telescope operated on the Spanish Observatorio del Teide on the island of Tenerife by the Aarhus and Copenhagen Universities and by the Instituto de Astrof\'{i}sica de Canarias.}}
\author{A. Bonanno\inst{1}\and
E. Corsaro \inst{1}\and
F. Del Sordo \inst{2,3} \and
P. L. Pall\'{e} \inst{4} \and
D. Stello \inst{5} \and
M. Hon \inst{5}
         }
\offprints{Alfio Bonanno\\ \email{alfio.bonanno@inaf.it}}

\institute{
INAF -- Osservatorio Astrofisico di Catania, via S. Sofia, 78, 95123 Catania, Italy
\and
    Institute of Astrophysics, FORTH, GR-71110 Heraklion, Greece
\and
    Department of Physics, University of Crete, GR-70013 Heraklion, Greece
\and
Instituto de Astrof\'{i}sica de Canarias, 38205 La Laguna, Tenerife, Spain
\and
School of Physics, University of New South Wales, NSW 2052, Australia
}

%A\&A 579, A83 (2015), DOI: 10.1051/0004-6361/201525895
%This corrigendum also includes modifications to:
%A\&A 578, A76 (2015), DOI: 10.1051/0004-6361/201525922
%
\abstract
{We present further evidence of the presence of acoustic oscillations on the slowly-rotating, over-active G8 sub-giant
EK Eri. This star was observed with the 1-m Hertzsprung SONG telescope, at the Observatorio del Teide for two different runs of 8 and 13 nights, respectively, and separated by about a year. 
We determined a significant excess of power around $\numax = 253 \pm 3\,\mu$Hz in the first observing run and we were able to determine the large separation $\dnu = 16.43 \pm 0.22\,\mu$Hz. No significant excess of power was instead detected in a subsequent SONG observing season, as also supported by our analysis of the simultaneous TESS photometric observations. 
We propose a new amplitude-luminosity relation in order to account for the missing power in the power spectrum. Based on the evolutionary stage of this object we argue that standard $\alpha^2\Omega$ dynamo cannot be excluded as a possible origin for the observed magnetic field.}

% 5 {} token are mandatory
%
%\abstract
%{ }
\titlerunning{Asteroseismology and dynamo action on EK Eridani}
      \authorrunning{A. Bonanno et al.}
\maketitle

%
%===================================================================================
\section{Introduction}
Although with different physical emphasis, both magnetic fields and acoustic oscillations are ubiquitous along the HR diagram.

In the case of the Sun the level of activity is correlated with the frequencies and power of the p-mode oscillations \citep{woodard85,Palle89,libbrecht90,Elsworth90}. 
In recent times, thanks to the high-quality data of the CoRoT and {\it Kepler} missions, it has been possible to investigate the effect of magnetic activity in solar-like stars in a systematic fashion \citep[e.g. see][]{Garcia10Science}. In \cite{Chaplin11activity} in particular, a sample of about 2000 solar-like stars has been analyzed and the evidence of pulsations was found only in about $\sim 540$ stars. In particular it was found that as the level of activity increased, the number of detections decreased. A further analysis that made use of the Mt. Wilson S-index chromospheric activity indicator, has demonstrated the presence of an anti-correlation between the S-index and the amplitude of the acoustic modes \citep{Bonanno14}. A possible explanation is the decrease of the turbulent velocities near the surface due to the presence of the magnetic field that reduces the efficiency of the convection \citep{dziembowski05, Jacoutot08}.

The relevant question is to understand the physical mechanism underlying this process. From this point of view the G8 sub-giant\footnote{We note that although the star is a spectroscopically determined sub-giant, its characteristic oscillation frequency, $\numax$, puts it firmly on the red giant branch.} EK Eri (HR 1362, HD 27536) represents a unique opportunity. As first noticed by \cite{Dall10}, analysis of high-precision radial velocities obtained with the HARPS spectrograph has revealed a significant suppression of the oscillation modes. 

Because of its slow-rotation $(v\sin i < 1$ $ {\rm km\,s^{-1}})$, brightness variations are consistent with a rotational period of $ \sim 308$ days due to star spots being rotated across the projected surface of the star \citep{strassmeier90,strassmeier99}.
Spectropolarimetric observations \citep{Auriere08EKEri,Auriere11} have determined that the magnetic field geometry is dominated by a large-scale almost completely dipolar field of strength $\sim 200$-$250$\,G. This object is therefore significantly over-active with respect to its rotation rate and evolutionary state, being severely off the normal period-activity relation \citep{auriere15}.

The physical explanation for the presence of the observed field in this object is still much debated. Assuming conservation of the magnetic flux, EK Eri on the ZAMS would have had a field strength of a few kG, which is well within the typical range for a magnetic Ap star \citep{stepien93}. However, this simple assumption is almost certainly inadequate because, as the star evolves off the main sequence, the growing outer convection zone interacts with the original field.
Therefore, a combination of a fossil field acting as the seed for a growing dynamo generated field may have emerged. However, the abundance pattern determined by \cite{Dall10} is very similar to the solar one, suggesting that there are no anomalies that could be attributed to the previous evolutionary status as a magnetic Ap star. 

Alternatively, in spite of its very low equatorial velocity, EK Eri might have produced enough differential rotation to host a genuine dynamo that generated the observed field. In fact, a recent investigation with \textit{Kepler} data has shown that, in contrast to the well established behavior at small Rossby numbers (the ratio of rotation period to the convective turnover time), the chromospheric activity of the more slowly rotating stars of  the open cluster M67 has been found to increase with increasing Rossby number \citep{axel18}.  
According to \cite{axel18}, slow-rotators with enhanced stellar activity might indeed be characterized by anti-solar differential rotation.

The study of the acoustic spectrum of EK Eri will provide essential information to settle the question. Asteroseismology can precisely determine the evolutionary status of the star and therefore constrain its internal density stratification and determine the strength of the core-envelope coupling from the spectrum of mixed-modes. However, the severe violation of the scaling relation for the amplitude \citep{kjeldsen95} when comparing to the value measured by \cite{Dall10}, if properly interpreted, opens up a window into the details of the excitation mechanism of the acoustic modes in the presence of a strong surface magnetic field. 

In this work new radial-velocity measurements obtained with the SONG spectrograph obtained in two different observational campaigns will provide further evidence for the presence of  $p$-modes in this star. Albeit strongly suppressed the location of the excess of power in the power spectrum allowed us to obtain an estimation for the large separation on this star. 

We therefore used this information to infer an internal model of the star and we use a spherical dynamo model to study possible
dynamo actions in this object. We show that an $\alpha^2\Omega$ dynamo can be operative even if the Rossby number near the base
of the convection zone is greater than one. The resulting magnetic field topology is close to the observed one. 

We further suggests a modification of the standard amplitude scaling relation of \cite{kjeldsen95} that takes into account the missing convective flux due the presence of a large spotted area. We argue that the non-detection of the oscillation modes in the second observational campaign is a consequence of further decrease of the efficiency of convection due to the large fluctuations of the magnetic field, as observed in \cite{Auriere11}.

The structure of the paper is as follows: in {Sect.~2} we describe the observations and data reduction, {Sect.~3} describes the results, Sect.~4  contains the results of a kinematic dynamo-model applied to the star, {Sect.~5} is devoted to the conclusions. 

\section{Observations}
\label{sec:obs}
\cite{Dall10} reported the first detection of solar-like oscillations in EK Eri, using about three nights of high-precision radial velocity observations with HARPS. However, the short length of their observation set allowed for estimating only the frequency of maximum oscillation power, $\numax$, which they found to be $320\pm32\,\mu$Hz. Following on this detection, we decided to perform new observations of the target to be able to better characterize its $\numax$ and possibly derive its large frequency separation $\dnu$ for the first time.

\begin{figure}[h]
   \centering
   \includegraphics[width=0.48\textwidth]{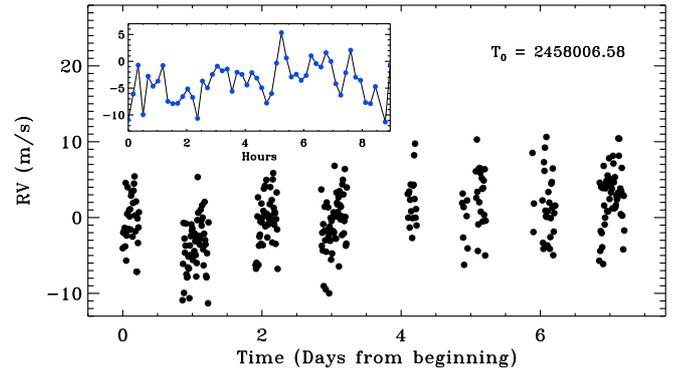}
\caption{Radial velocity time-series of EK Eri with SONG for the entire 2017 run, with the detail of the second night of observations shown in the inset. The time $T_0$ (BJD) of the first data point of the entire campaign is also indicated.}
         \label{fig:ts}
   \end{figure}

The new radial velocity time-series obtained for EK Eri come from the observations with the 1-m Hertzsprung SONG telescope, at the Observatorio del Teide, using its e\'chelle spectrograph. The Hertzsprung telescope constitutes the first node of the Stellar Observations Network Group (SONG). Observations were conducted for a total of about eight consecutive nights starting from December 2$^{\rm nd}$ 2017, and with $\sim$312 individual data points, so about 40 per night (see Fig.~\ref{fig:ts}). The spectra were reduced and calibrated using the SONG pipeline and an iodine cell for precise wavelength calibration. For more details about the Hertzsprung telescope characteristics and reduction pipeline we refer to \citet{Andersen14,Grundahl17}. Given that the star has an apparent magnitude of $V \simeq 6.1$, it was chosen to use a spectral resolution of 90\,000 and an exposure time of 600\,s throughout the run. A second observing run of $\sim13$ nearly consecutive nights was also performed during 2018, and we discuss its results in Sect.~\ref{sec:second_ts}.

\subsection{Time-series analysis}
\label{sec:ts_analysis}
%
%______________________________________________________________

The power spectrum of the velocity time-series was calculated as both an un-weighted and weighted least-squares fit of sinusoidal components following \cite{Corsaro12} \citep[see also][]{Frandsen95,Arentoft98,Bedding04,Kjeldsen05,Creevey19}. The resulting power was then converted into power spectral density (PSD) by normalizing for the spectral resolution, which is calculated as the integral of the spectral window of the observing set, here corresponding to $\sim4.87\,\mu$Hz. The purpose of testing the case of a weighted least-squares spectrum is to check for any improvements in the signal-to-noise ratio, which is relevant in order to increase the chances of detecting an oscillation envelope. Similarly to what has been done already in \cite{Corsaro12}, we assigned a weight to each point of the radial velocity time-series that was given by the corresponding uncertainty estimate from the SONG pipeline. Since radial velocity uncertainties provided by the SONG pipeline might not necessarily be realistic, we have corrected them for possible outliers following the approach presented by \cite{Butler04} and adopted by \cite{Corsaro12}. By measuring the amplitude of the noise level in the amplitude spectrum in the region $300\,\mu$Hz up to the Nyquist frequency of $\sim822\,\mu$Hz, we found that the amplitude of the noise is $34.0$\,cm\,s$^{-1}$ and the maximum amplitude around $\numax$ is $99.8$\,cm\,s$^{-1}$ in the un-weighted case, while we have $33.5$\,cm s$^{-1}$ and $106.0$\,cm\,s$^{-1}$, respectively, in the weighted case. The signal-to-noise ratio is slightly improved in the weighted spectrum, hence we adopted this latter one for the subsequent asteroseismic analysis.

%----------------------------------------------------------- S_vib
   \begin{figure}[h]
   \centering
   \includegraphics[width=0.5\textwidth]
{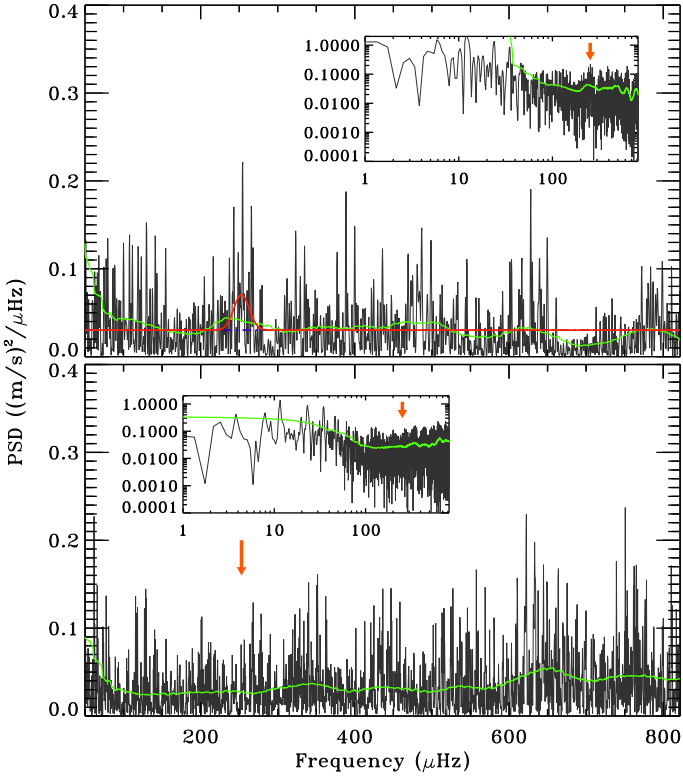}
\caption{\textit{Top panel}: Power spectral density (in black) and background model fit using \diamonds\,\,for EK Eri from the 8-nights observation run in 2017. The red line shows the total fit including the oscillation power excess, while the dashed blue line below the oscillation hump represents the flat noise component. The inset shows the same spectrum but in a log-log scale to help visualizing the hump due to oscillations, which is also marked by a red arrow at $\numax$. A smoothing by 4$\dnu$, with $\dnu$ derived from \cite{Stello09} using $\numax$ from our analysis (solid green line), is also shown. \textit{Bottom panel}: Similar to the top panel but here using the second observation run from 2018. The position of $\numax$ from the 2017 time-series is indicated by the red arrows, showing that there is no region where significant power excess is found.}
         \label{fig:psd_fit}
   \end{figure}
%
%______________________________________________________________

%----------------------------------------------------------- S_vib
   \begin{figure}[h]
   \centering
   \includegraphics[width=0.5\textwidth]
{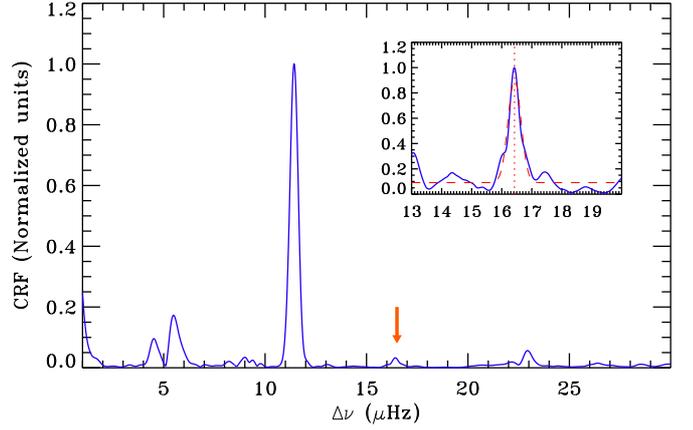}
\caption{The CRF of the region of the PSD of EK Eri centered around $\numax$ (in blue) for a search range between 1 and 30\,$\mu$Hz, where multiples of the daily aliasing at 11.57\,$\mu$Hz are clearly visible. The arrow indicates the peak corresponding to the position of the actual $\dnu$. The inset shows the result of applying the CRF to a smaller frequency range that excludes the main peaks from the daily aliases, between 13 and 20\,$\mu$Hz. In the inset, the red dashed curve represents a Gaussian fit to the CRF peak, with the centroid and standard deviation of the Gaussian marking the value of $\dnu$ and its corresponding uncertainty, as used in this analysis.}
         \label{fig:dnu}
   \end{figure}
%
%______________________________________________________________

\subsection{Extraction of global seismic parameters}
Similarly to \cite{Creevey19} we used the \diamonds\,\,code \citep{Corsaro14} for the Bayesian inference of the PSD of EK Eri. The model adopted consists in a simple flat noise component and a Gaussian envelope to model the oscillation power excess \citep{Corsaro14}. To identify the power excess, we searched in a region close to the first $\numax$ estimate published by \cite{Dall10}. This region can also be identified by means of the spectroscopic properties of $\log g$ and $T_\mathrm{eff}$ obtained by \cite{Dall10}, given that $\numax \propto g/\sqrt{T_\mathrm{eff}}$ \citep{brown11}. Using the new estimate for $\nu_\mathrm{max,\odot}$ by \cite{SONGSun} to scale from the Sun, the spectroscopic values yield $\numax \approx 300\,\mu$Hz. This is well below the limiting Nyquist frequency of our dataset. The resulting fit with \diamonds\,\,is shown in Fig.~\ref{fig:psd_fit} (top panel) where the Gaussian-shape hump of the oscillations is centered at $\numax = 253 \pm 3\,\mu$Hz, with the uncertainty corresponding to the 68.3\,\% Bayesian credible interval (see Table~\ref{tab:prop} for a list of all the properties of this star). We additionally tested the significance of this region by means of a statistical model comparison against: 1) a non-detection case (only flat noise); 2) the selection of a different power hump, one at 330\,$\mu$Hz and another around 130\,$\mu$Hz, which represent the two closest neighbours to the one originally selected. In all cases investigated, the result based on confronting the Bayesian evidences of each model (with and without the Gaussian envelope and between the different power humps selected), yields a strong evidence condition ($\ln B > 5$, where $B$ is the Bayes' factor) in favor of the detection of the oscillation envelope centered at $\numax = 253\,\mu$Hz. 

In addition to a statistical validation of the power excess, we have estimated the characteristic frequency of the granulation component, $b_\mathrm{gran}$, using the photometric TESS observations of EK Eri (see Sect.~\ref{sec:tess} for more details). This approach provides a totally independent check on the expected value of $\numax$ for our star because $b_\mathrm{gran}$ and $\numax$ are tightly related to one another \citep[see e.g.][]{Kallinger14}. By performing a background fit with \diamonds\,\,to the TESS PSD \citep{Corsaro17}, we have that $b_\mathrm{gran} =  230$\,$\mu$Hz. By inverting the power-law linking $b_\mathrm{gran}$ to $\numax$ as obtained by \cite{Kallinger14}, we have $\nu_\mathrm{max,pred} = 253$\,$\mu$Hz, which is therefore in excellent agreement with our estimate from the SONG spectrum.

For determining the value of $\dnu$ we instead focused on the region of the oscillations and computed a comb response function (CRF) out of a set of the 10 highest SNR frequencies extracted by means of a CLEAN algorithm in the PSD region 200-300\,$\mu$Hz. For this purpose we adopted a similar procedure as that presented in \cite{Bonanno08} and later on applied by \cite{Corsaro12}. We first selected a wide range of values for $\Delta\nu$ ($1$-$30$\,$\mu$Hz) in order to clearly locate the presence of a possible peak different than the multiples of the daily aliasing, $11.57$\,$\mu$Hz. Then, we restricted our search range to a region where $\dnu$ from scaling relations is expected (e.g. see \citealt{Stello09}), hence excluding the daily aliases at $11.57$ and twice this value in order to enhance the signal of the expected $\dnu$ peak. The result is presented in Fig.~\ref{fig:dnu}, where we can observe a peak of the CRF at $\dnu = 16.43 \pm 0.22\,\mu$Hz.

Lastly, we also determined the maximum amplitude of the radial oscillations, $A_{\rm osc}$, following the recipe presented by \cite{Kjeldsen05} (but see also \citealt{Corsaro12}), where the uncertainty reported in Table~\ref{tab:prop} is obtained by incorporating the uncertainty on the background level of the PSD.

\section{Results}
\label{sec:results}
\subsection{Derivation of fundamental stellar properties}
Because a Gaia DR2 parallax is available \citep{gaia2018b}, we first compute a stellar radius based on this information, exploiting the interstellar reddening from the PanSTARRS-1 3D Galaxy map \citep{Green15dust3D}, the extinction/reddening ratio for high galactic latitude stars from \cite{Fitzpatrick99reddening}, and a bolometric correction from \cite{Flower96} for a $T_{\rm eff} = 5135\,$K star \citep{Dall10}. The result obtained is $R_{\rm GDR2} = 4.91 \pm 0.13\,R_{\rm \odot}$. We note that although the Gaia DR2 parallax is not within 1-$\sigma$ of the Hipparcos parallax for the same star, we still find a good agreement (well within 1-$\sigma$) between our estimate of stellar radius based on the Gaia DR2 parallax and that obtained by \cite{Dall10} using the Hipparcos parallax.

Once the global seismic parameters are obtained, we can obtain an estimate of stellar mass and radius from asteroseismic scaling relations \citep[e.g.][]{kjeldsen95}, using the relations
\begin{equation}
M/M_{\odot} = \left(\frac{\numax}{\nu_{\rm{max,\odot}}} \right)^3 \left( \frac{\dnu}{\dnu_{\odot}}\right)^{-4} \left( \frac{T_{\rm eff}}{T_{\rm eff,\odot}}\right)^{1.5}
\end{equation}
\begin{equation}
 R/R_{\odot} = \left(\frac{\numax}{\nu_{\rm{max,\odot}}} \right) \left( \frac{\dnu}{\dnu_{\odot}}\right)^{-2} \left( \frac{T_{\rm eff}}{T_{\rm eff,\odot}}\right)^{0.5}
\end{equation}
for mass and radius, respectively, where we adopted $T_{\rm eff, \odot} = 5777$\,K and the new asteroseismic Solar-SONG reference values \citep{SONGSun}, namely $\nu_{\rm max, \odot} = 3141 \pm 12\,\mu$Hz, $\dnu_{\odot} = 134.98 \pm 0.04\,\mu$Hz. We note that the seismic radius obtained in this work, $R_{\rm seismic} = 5.12 \pm 0.15\,R_{\rm \odot}$ is  compatible (within 1-$\sigma$) with that derived using the Gaia DR2 parallax, and having a precision of about 3\,\%. We derive a seismic mass of $M_{\rm seismic} = 2.00 \pm 0.14\,M_{\rm \odot}$, which with a precision of about 7\,\%, is in agreement with previous estimates proposed by \cite{Auriere08EKEri}, and in particular with the measurement obtained by \cite{Dall10} (well within 1-$\sigma$). We also note that this value is in agreement with a mass derived using the spectroscopic $\log g$ from \cite{Dall10} and the new stellar radius obtained from Gaia DR2, corresponding to about $2.16\,M_{\odot}$. However, we point out that standard (uncorrected) seismic scaling relations have the tendency to overestimate stellar radius and mass when applied to evolved stars \citep[see e.g.][]{Gaulme16scaling,Sharma16}. For mitigating this problem, \cite{Kallinger18nonlinear} recently developed new non-linear scaling relations. In order to check for possible inconsistencies, we have adopted these non-linear relations for EK Eri, obtaining $R_\mathrm{non-linear} = 4.96 \pm 0.14\,R_{\rm \odot}$, $M_\mathrm{non-linear} = 1.84 \pm 0.12\,R_{\rm \odot}$. These values, although systematically lower, appear to be still in agreement (within 1-$\sigma$) with radius and mass estimates obtained from the standard scaling relations adopted in this work. The non-linear scaling estimates are also compatible (within 1-$\sigma$) to literature values from \cite{Auriere08EKEri,Dall10} and to the Gaia DR2 radius and logg-based mass.

By exploiting the new Gaia parallax to derive the stellar luminosity, we find that this is in agreement with that derived by \cite{Dall10}, but with the advantage that in this work it  is estimated with a precision of about 2.3\,\%, as opposed to a precision of about 11\,\% of the previous value.
%__________________________________________________ One column table
   \begin{table}
      \caption[]{Observed Properties of EK Eri used in this work.}
         \label{tab:prop}
         \footnotesize
         \begin{tabular}{llll}
            \hline\hline\\[-8pt]
            Property & Units & {\rm Value} & {\rm Source}      \\[1pt]
            \hline\\[-8pt]
			$v \sin i$ & km\,s$^{-1}$ & $< 1.6 \pm 0.4$ & \citet{Dall10}\\[1pt]
            $P_{\rm rot}$ & d & $308.8 \pm 2.5$ & \citet{Dall10}\\[1pt]
            $T_{\rm eff}$ & ${\rm K}$ & $5135 \pm 60$ & \citet{Dall10}\\[1pt]
            $\log g$ & & $3.39 \pm 0.12$ & \citet{Dall10}\\[1pt]
            $\left[\mbox{Fe/H}\right]$ & dex & $+0.02\pm 0.04$& \citet{Dall10}\\[1pt]
            \hline\\[-8pt]
			$E(B-V)$ & & 0.006 & \citet{Green15dust3D}\\[1pt]
			$A_v/E(B-V)$ & & 3.1 & \citet{Fitzpatrick99reddening}\\[1pt]
        	$BC$ & & $-0.25$ & \citet{Flower96}\\[1pt]
        	$\pi_{\rm GDR2}$ & $\mbox{mas}$ & $15.58 \pm 0.05$ & \citet{gaia2018b}\\[1pt]
            $R_{\rm GDR2}$ & $R_{\odot}$ & $4.91 \pm 0.13$ & This work\\[1pt]
            $L$ & $L_{\odot}$ & $15.07 \pm 0.35$ & This work\\[1pt]
            \hline\\[-8pt]
            $\numax$ & $\mu$Hz & $253 \pm 3$ & This work\\[1pt]
			$\dnu$ & $\mu$Hz & $16.43\pm 0.22$ & This work\\[1pt]
			$A_{\rm osc}$ & m\,s$^{-1}$ & $0.22 \pm 0.01$ & This work\\[1pt]
            $R_{\rm seismic}$ & $R_{\odot}$ & $5.12 \pm 0.15$ & This work\\[1pt]
            $M_{\rm seismic}$ & $M_{\odot}$ & $2.00 \pm 0.14$ & This work\\[1pt]
            \hline\hline
         \end{tabular}
   \end{table}
%
%% %
% %______________________________________________________________

\subsection{Suppression of the oscillation amplitude}
As shown by e.g. \cite{Garcia10Science,Chaplin11activity}, the presence of surface magnetic activity has the effect of suppressing the amplitude of solar-like oscillations as the result of a less efficient convection mechanism.  By considering the scaling relation for oscillation amplitudes in radial velocity, proposed originally by \cite{kjeldsen95}, namely
\begin{equation}
\label{kb95}
A_{\rm osc} = \frac{L/L_{\odot}}{M/M_{\odot}} (23.4 \pm 1.4)\,\,\mbox{cm\,s}^{-1}    \, ,
\end{equation}
and based on our derivation of the stellar luminosity and mass, we evaluate that the expected oscillation amplitude 
for EK Eri would be $A_{\rm osc} = 1.79 \pm 0.11$\,m\,s$^{-1}$. 
This value is clearly significantly higher than what we measured from the actual PSD, by a factor of about eight,
while it appears close to the value measured by \cite{Dall10}, although about 30\,\% larger. 

We can interpret this fact in the following way.  We can imagine that if the star has a large spotted area, the effective luminosity responsible for the amplitude in Eq.~(\ref{kb95}) is not $L$ but
$L-S\sigma T^4$ where $S$ is the area of the spot and 
$\sigma$ is the Stefan-Boltzmann constant. Therefore Eq.~(\ref{kb95}) can be rewritten as
\begin{equation}
\label{newscaling}
A_{\rm osc}^{\rm spotted} = \beta \frac{L/L_{\odot}}{M/M_{\odot}} (23.4 \pm 1.4)\,\,\mbox{cm s}^{-1}    \, ,
\end{equation}
where
\be
\beta = \frac{L-\Delta L}{L} = 1 - \frac{S}{4 \pi R^2}
\label{eq:spotted_factor}
\ee
given that $\Delta L = S \sigma T^4$.
If we use our information about the suppression factor of the p-mode amplitude, we can easily invert Eq.~(\ref{newscaling})
and estimate the effective spotted area, which turns out to be $88\,\pm 6$\,\% of the total surface of the star. It is encouraging
to notice that this estimate is roughly in agreement with the findings of \cite{Auriere11} (see in particular the topology of the radial
field in their Fig. 6). Clearly for a much less active star like the Sun, $\beta\approx 1$ and one recovers the standard scaling relation. 

We note that when using a {\it Kepler} calibrated amplitude scaling relation as the one proposed by \cite{Corsaro13} (converted into a radial velocity amplitude according to the relation by \citealt{kjeldsen95}) we obtain $A_{\rm osc} = 1.24$\,m\,s$^{-1}$,which is 30\,\% smaller than that predicted using Eq.~(\ref{kb95}). Relative to this expected oscillation amplitude the observed amplitude suppression is still a factor of almost six, and the spotted area estimated through Eq.~(\ref{eq:spotted_factor}) remains about 80\,\% of the total visible surface of the star. However, one should keep in mind that the conversion from bolometric to radial velocity amplitude following \cite{kjeldsen95} is still not fully understood and that the derived $A_{\rm osc}$ in this latter case might be overestimated \citep[see e.g.][]{Huber11Procyon}.

\subsection{About the missing oscillatory signal in the second SONG time-series}
\label{sec:second_ts}
\begin{figure}[h]
   \centering
   \includegraphics[width=0.48\textwidth]
{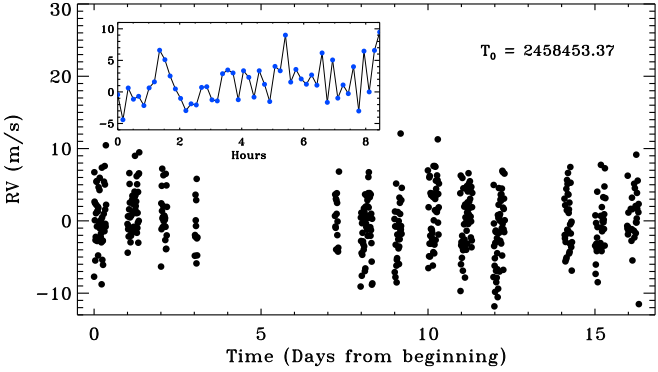}
\caption{Radial velocity time-series of EK Eri with SONG for the entire 2018 run, with the detail of the second night of observations shown in the inset. The time $T_0$ (BJD) of the first data point of the entire campaign is also indicated. No clear periodicity in the signal appears visible.}
         \label{fig:ts_2018}
   \end{figure}

In addition to the first observations of EK Eri that we presented in Sect.~\ref{sec:obs}, we attempted to observe the star one year later, using the same SONG instrumental setup, in order to gather a longer time-series and thus obtain a better frequency resolution on the resulting PSD. The new observations were conducted between November 30$^{\rm th}$ 2018 and December 17$^{\rm th}$ 2018, covering a total of about 13 nights and nearly 480 individual RV measurements (see Fig.~\ref{fig:ts_2018} for more details on the distribution of individual nights of observation).

Observing conditions were very similar as during the first observations in 2017, and so was the noise level measured in the data, which is found to be $31.5$\,cm\,s$^{-1}$ as measured in the region from 300\,$\mu$Hz up to the Nyquist frequency. Nonetheless, by processing the time-series in the same manner as presented in Sect.~\ref{sec:ts_analysis}, we do not find any evidence for a power excess due to the oscillations in the frequency region of $\numax$ from Table~\ref{tab:prop}, as it appears visible from the new PSD of the star (see Fig.~\ref{fig:psd_fit}, bottom panel). Since the noise level of the new dataset is not higher than the one from 2017, we can exclude the noise in the PSD as the possible cause of a non detection. In this regard, \cite{Auriere11} noted that the magnetic field is subject to fluctuations on the order of 80\,G from one year to another (see also their Table~1). We deem such fluctuations as the likely cause of a variation in the level of suppression of the oscillations and therefore of the missing oscillation power excess. According to the previous discussion, even a small increase of $\beta$ can render the amplitude
of the modes not observable given the already small signal to noise level.

\subsection{New TESS observations}
\label{sec:tess}
\begin{figure}[h]
   \centering
   \includegraphics[width=0.48\textwidth]{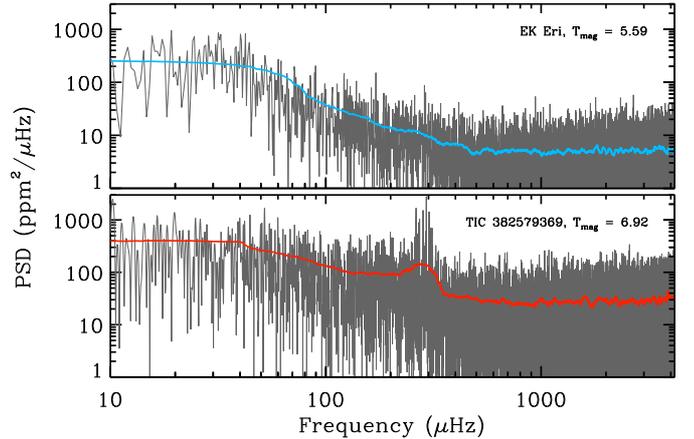}
\caption{Upper panel: PSD (in gray) for EK Eri using the TESS observations from sector 5. Lower panel: the PSD of TIC~382579369, a star with similar magnitude and $\numax$ of EK Eri. The colored solid lines represent the smoothing by 4$\dnu$, similarly to what is done in Fig.~\ref{fig:psd_fit}.}
         \label{fig:tess}
\end{figure}

From 15 November till 11 December 2018, EK Eri (TIC~37778297) was monitored by the recently launched NASA TESS mission \citep{Ricker14TESS} during observing sector 5. This photometric observation is essentially simultaneous to that of our second SONG time-series presented in Sect.~\ref{sec:second_ts}, thus providing us with the possibility to assess the detectability of the oscillation envelope from both photometric and radial velocity datasets. To further aid this assessment, we also analysed the TESS data of a star with similar $\numax$ and brightness as EK Eri that would serve as a comparison star. We applied 4-sigma clipping to the TESS light curves and passed them through a high-pass filter (9.3-hour wide boxcar) with a cut-off frequency of 30\,$\mu$Hz to remove long-term trends. Finally, the light curves were gap-filled using the in-painting algorithm \citep{Pires15} to reduce the effect of spectral leakage of high power from low frequencies that could otherwise obscure the power-frequency profile of solar-like oscillations. The resulting power spectrum of EK Eri as observed with TESS is shown in Fig.~\ref{fig:tess} (upper panel), where we also plot the TESS comparison star, TIC~382579369 (lower panel). The comparison between the two shows that the lack of a clear oscillation power excess in EK Eri is not the result of the star being too faint (the comparison star is in fact slightly fainter), but rather that EK Eri shows strongly suppressed oscillations. The absence of an oscillation bump in the TESS power spectrum of EK Eri therefore supports the conclusions drawn from the analysis of our second SONG time-series.

\begin{figure}[h]
   \centering
   \includegraphics[width=0.48\textwidth]{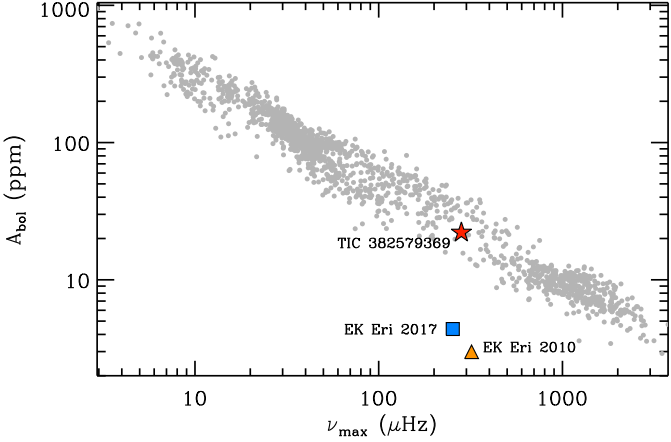}
\caption{The bolometric oscillation amplitudes for a sample of 1640 \textit{Kepler} stars (gray dots) analyzed by \cite{Huber11}. The red star marks the amplitude of the comparison star TIC~382579369 (Fig.~\ref{fig:tess}, lower panel) as measured in this work, while the blue square shows the amplitude of EK Eri obtained by converting its radial velocity amplitude into a bolometric one. The earlier measurement by \cite{Dall10} is also included using a yellow triangle.}
         \label{fig:amp_numax}
\end{figure}

To further quantify the level of suppression, we also perform a background fit to the power density spectrum of TIC~382579369 with \diamonds, using the same approach adopted by \cite{Corsaro15,Corsaro17}. In this way, we can estimate the photometric amplitude of our TESS comparison star following the same procedure used by \cite{kjeldsen95}. We show the result in Fig.~\ref{fig:amp_numax}, where the amplitude of TIC~382579369, of $\sim$22\,ppm (star symbol), is compared to a sample of \textit{Kepler} stars ranging from the main sequence (high $\numax$) to near the tip of the red giant branch (low $\numax$) that were previously analyzed by \cite{Huber11} using the same method (see also \citealt{Corsaro13}). In the same plot, we also show our estimated bolometric amplitude for EK Eri, 4.38\,ppm, as obtained by converting the radial velocity amplitude from the SONG 2017 observing campaign into a bolometric one according to \cite{kjeldsen95}. It is clear that while the amplitude of TIC~382579369 is in line with the `normal' trend from the \textit{Kepler} sample, EK Eri falls significantly below what is expected for a star having $\numax$ of about 200-300\,$\mu$Hz.

\section{A possible dynamo explanation}
It is interesting to investigate the possibility that an $\alpha^2$ or $\alpha\Omega$ dynamo mechanism is at work in this star \citep{krause}. 
In the standard $\alpha\Omega$ dynamo \citep{parker55}, the combined action of shear and mirror-antisymmetric turbulence regenerates the initial poloidal field and closes the dynamo loop. The field configurations are in general rather complex, and the solutions are oscillatory. On the contrary, if the shear is not dominant, the rotational  helicity of the turbulence alone amplifies the seed field up to the observed values. The resulting field is in general static and the geometry rather simple, mostly  dipolar.
Only the turbulent diffusivity $\eta$ can be non-zero for a non-rotating isotropic turbulence. In the Sun it is believed that both mechanism are at work \citep{bo13,gustavo16}, although the effect of meridional circulation (together with a low-eddy diffusivity) is essential in order to reproduce the observed butterfly diagram.

The astrophysical parameters obtained in the previous sections provide us with important information about the evolutionary state of the star, in particular 
the extension of the convective zone and the efficiency of the turbulence. To this purpose we used the Catania version of the \gar code 
\citep{garstec,bosch02} to determine its interior structure. In particular  we  evolved from the ZAMS a $2 M_\odot$   model with initial He abundance and metal
fraction $Y_i=$0.3, and $Z/X=0.045$, respectively, up to a final age of $0.965$ Gyr in order to match the observed radius and luminosity. No heavy-element diffusion has been included and a mixing-length parameter $\alpha_{\rm MLT}=1.65$, as obtained from solar calibration, has been chosen.

\begin{figure}[h]
	\centering
\includegraphics[width=0.5\textwidth]{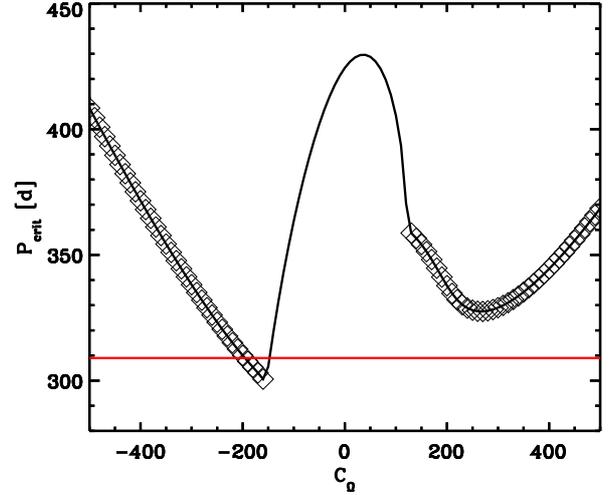}
\caption{Critical period as a function of $C_\Omega$, the critical dynamo number of the differential rotation. Negative $C_\Omega$ indicate that the angular velocity
decreases outwards. The diamond represent oscillatory solutions. The cycles periods are of the order of $50$ years. The red line is the observed period.}
	\label{fig:pcrit}
 \end{figure}
   
   \begin{figure}[h]
   \centering
   \includegraphics[width=0.4\textwidth]{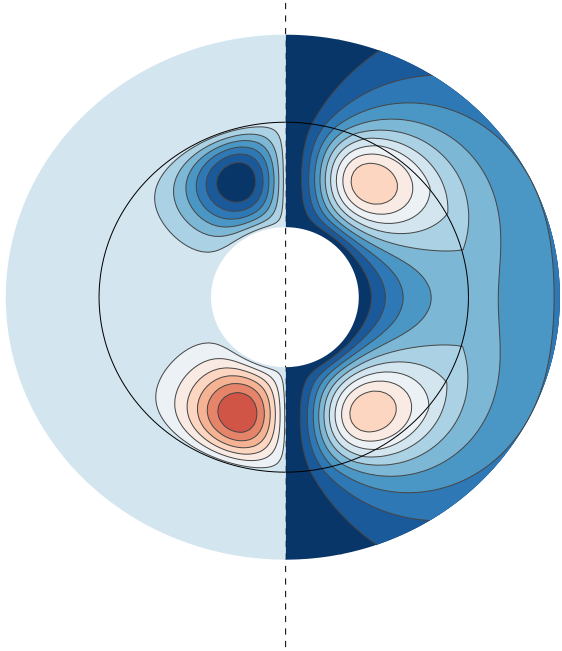}
\caption{Topology of the field for $C_\Omega=50$ (see text). The left hemisphere represents the isocontour lines of the toroidal field
with blue levels for negative toroidal field and red for positive values. 
The right hemisphere represents the streamlines of the poloidal field. Blue levels are for
counterclockwise field lines; red levels for clockwise field lines. The black circle is the surface
of the star.
}
         \label{fig:topology}
   \end{figure}

The relative importance of rotation and turbulence is described by the  Coriolis  number $\Omega^\ast = 2 \Omega \tau$ where $\tau$ is the convective turnover time and $\Omega$ the surface angular velocity. In our case it is of order of unity in the bulk of the convection zone and near the surface $\Omega^\ast \ll1$, as expected for such a slow rotator. The anti-symmetric component of turbulent electromotive force can be approximated with a pseudo-scalar of the type $\alpha(r)\cos(\theta)$. In particular for small Coriolis number  \citep{RK93,KuR99}   
$\alpha \approx \Omega^\ast \nabla \log(\rho) \tau u^2 
\approx \Omega \ell^2/R$ where $u$ is the (turbulent) convective velocity field and 
$\ell\approx 0.20 R$ is the characteristic radial scale of the turbulence.

The equipartition field $B_{eq} = \sqrt{4\pi \rho} u$ turns out to be  of order $10^3$ G at the bottom of the convection zone ($r_{b}=0.4R$) and it reaches $\sim 100$ G 
near the surface. 
Although this value is of the order of the observed one it would be difficult to explain the generation of  a large, 
long-lasting spot without the presence of a large-scale field  produced by a suitable dynamo action. 
We can model our ignorance on the differential rotation by assuming a simple radial law of the type $\Omega = \Omega_0 (r-r_i)/R$ where $\Omega_0$ is the observed
equatorial angular velocity. With  this definition, with a turbulent (eddy) diffusivity of the order of   $\eta\sim 10^{14} {\rm cm^2 s^{-1}}$
(asssuming the mixing-length relation $\eta = u^2 \ell /3$ ), the dynamo number of the shear $C_\Omega = \Omega_0 R^2/\eta$ is around $\sim 250-300 $.

Is the rotational period of EK Eridani short enough to support a dynamo action? A possible answer to this question can be obtained within the mean-field framework by solving the turbulent induction equation and determine the critical value of $\alpha$ necessary to obtain a self-sustained large scale field.
In particular, we used the \ctd mean field dynamo code presented in  \cite{boel02} and further discussed and tested in \cite{jouve10}. The results are depicted in Fig.(\ref{fig:pcrit}). We compute the critical period $P_{\rm crit}$ below which a dynamo action is possible, and this is plotted as a function of the rotational shear  $C_\Omega$. Negative $C_\Omega$ implies that the angular velocity decreases in radius, whilst diamonds represent critical oscillatory solutions, and the cycle periods ranges from $50$ to $100$ yrs. A typical field configuration obtained with this approach is depicted in Fig.(\ref{fig:topology}), for $C_\Omega=50$.

\section{Conclusions}
EK Eri is a unique laboratory to study the behavior of acoustic pulsations in the presence of a large-scale magnetic field. 
This is clearly an important issue that at the moment is far from a complete understanding.

From the observational point of view, we confirm the $p$-modes detection already discussed in \cite{Dall10}, and we obtain a tentative measurement
of the large separation for the first time in this star. We also propose to modify the standard amplitude-luminosity variation of 
\cite{kjeldsen95} in order to take into account the missing energy flux due to the presence of a large spot.

Concerning the origin of the field in this object, we observe that the combined effect of rotation and turbulence is slightly supercritical and the dynamo action is thus possible. 
Moreover a dynamo of the $\alpha^2\Omega$ type, or any dynamo characterized by an $\alpha$-tensor describing the standard mirror-antisymmetric turbulence, implies a dipolar-like magnetic field topology in the case of low values of $C_\Omega$. This is in agreement with observations that suggest a dipolar topology of the field.
Clearly our result is heavily dependent on the assumed differential rotation. An approach based on numerical simulations will hopefully clarify the question \citep{fab19}.

\begin{acknowledgements}
We would like to thank Tim Bedding for pointing out that the TESS data had become available for EK Eri while we were working on this paper. We thank the referee Thomas Kallinger for useful comments to our manuscript. E.C. is funded by the European Union'Äôs Horizon 2020 research and innovation program under the Marie Sklodowska-Curie grant agreement No. 664931. Based on observations made with the Hertzsprung SONG telescope operated on the Spanish Observatorio del Teide on the island of Tenerife by the Aarhus and Copenhagen Universities and by the Instituto de Astrof\'{i}sica de Canarias. The work of F.D.S. has been performed under the Project HPC-EUROPA3 (INFRAIA-2016-1-730897), with the support of the EC Research Innovation Action under the H2020 Programme; in particular, F.D.S. gratefully acknowledges the support and the hospitality of INAF Astrophysical Observatory of Catania.
\end{acknowledgements}

\bibliographystyle{aa}

\end{document}